# Outage Constrained Secrecy Rate Maximization Using Cooperative Jamming

Shuangyu Luo, Jiangyuan Li, Athina Petropulu
Rutgers University, Piscataway, NJ 08854

*Abstract*—[1] We consider a Gaussian MISO wiretap channel, where a multi-antenna source communicates with a single-antenna destination in the presence of a single-antenna eavesdropper. The communication is assisted by multi-antenna helpers that act as jammers to the eavesdropper. Each helper independently transmits noise, which lies in the null space of the helper-destination channel, thus creating no interference to the destination. Under source-eavesdropper channel uncertainty, we derive the optimal source covariance matrix that maximizes the secrecy rate subject to probability of outage and power constraints. Further, for the case in which the helper-eavesdropper channels follow a zero-mean Gaussian model with known covariances, we derive the outage probability in a closed form. Simulation results in support of the analysis are provided.

*Index Terms*—Gaussian MISO wiretap channel, cooperative jamming, artificial noise, outage probability

## I. INTRODUCTION

Physical layer secrecy exploits channel conditions to maximize the rate at which reliable information is delivered to the legitimate destination, with the eavesdropper being kept as ignorant of that information as possible. This line of work was pioneered by Wyner [1], who showed that when the source-eavesdropper channel is a degraded version of the source-destination channel, the source and the destination can exchange secure messages in perfect secrecy at a non-zero rate, while the eavesdropper can learn almost nothing about the messages based on its observations. The need for physical layer security in the context of wireless communications is motivated by challenges associated with classical cryptographic approaches, most notably the exchange and maintenance of private keys.

The secrecy capacity for multiple antenna wiretap channels with perfect channel state information on the eavesdropper is established in [2], [3], [4] under sum power constraints, and in [5] and [6] under power-covariance constraints. For MISO wiretap channels, the optimal input covariance matrix that achieves the secrecy capacity under a sum power constraint was given in closed form in [7]. When the channel to the destination encounters more fading than the channel to the eavesdropper, a positive secrecy rate is difficult to guarantee. One way to overcome this problem is to use helpers who amplify-and-forward, or decode-and-forward the source signal, or perform cooperative jamming (CJ). In the latter case, helpers do not need to receive nor relay the source message, but rather just transmit noise to degrade the channel to the eavesdropper and thus increase the secrecy rate. In [8], a multiple access wiretap channel was considered and it was shown that if the optimal power allocation policy does not allow a certain user to transmit, that particular user could increase the secrecy rate by transmitting artificial noise. In [9], multiple helpers were employed to transmit optimally weighted jamming signals that enforce nulling at the legitimate destination and maximize the secrecy rate. Subsequently, [10] obtained the optimal weights by avoiding the nulling at the destination, thus achieving higher secrecy rate.

In [11], a MISO wiretap system (i.e., a multi-antenna transmitter, a single-antenna legitimate receiver and a single-antenna eavesdropper) is studied. The transmitter constructs a Gaussian distributed artificial noise that lies in the null space of its channel to the legitimate receiver, and transmits a sum of signal and artificial noise. As no knowledge on the eavesdropper channel is assumed, the power of the artificial noise is designed to uniformly spread along the null space of the source-destination channel. In [12], the use of artificial interference for a MIMO wiretap channel is studied. The transmitter transmits a sum of signal and artificial noise. Again, no knowledge on the eavesdropper channel is assumed, and the artificial noise is designed to lie in the null space of the right singular vector associated with the largest singular value of the transmitter-receiver channel matrix. The power of the noise is uniformly spread along the null space. In [13], the artificial noise is studied in a MISO wiretap channel that includes multiple single-antenna eavesdroppers. The source transmits a mixture of message and artificial noise. Without the CSI of the eavesdroppers, the outage based artificial noise design is formulated, with a so-called safe convex approximation is used to find the solution.

In this paper, we consider a MISO wiretap channel with multiple multi-antenna helpers implementing cooperative jamming. While the multi-antenna source transmits the message, each helper transmits jamming noise that lies in a subspace that is orthogonal to the helper-destination channel. Each helper generates jamming noise locally, based on only its own link to the receiver. No knowledge on the helper-eavesdropper channel is assumed, thus the power of the jamming noise is uniformly spread along the helper-destination null space. We study the problem of determining the input covariance matrix that maximizes the secrecy rate subject to a sum power constraint and also an outage probability constraint. In order for the source to determine the optimal input covariance matrix only statistical information about the source-eavesdropper channel is required. A closed form expression for the outage probability is provides that applies to the case in which helper-eavesdropper channels follow zero-mean Gaussian distributions with known covariance matrix. Introducing an outage constraint not only provides quality control but also allows taking the uncertainty of the eavesdropper's channel into consideration and also simplifies the optimization problem with respect to the covariance matrix of the input.

*Notation* - Throughout this paper, following notation is adopted. Upper case and lower case bold symbols denote matrices and vectors, respectively. Superscripts $*$, $T$ and $\dagger$ denote respectively conjugate, transposition and conjugate transposition. $\text{Tr}(\mathbf{A})$ denotes the trace of the matrix $\mathbf{A}$. $\mathbf{A} \succeq 0$ denotes that the matrix $\mathbf{A}$ is Hermitian positive semi-definite. $|a|$ denotes absolute value of the complex number $a$. $\|\mathbf{a}\| = \sqrt{\mathbf{a}^\dagger \mathbf{a}}$ denotes Euclidean norm of the vector $\mathbf{a}$. $\mathbf{I}_n$ denotes the identity matrix of order $n$ (the subscript is dropped when the dimension is obvious). $\mathbb{C}^n$ denotes the set of all $n \times 1$ complex vectors. $\mathbb{E}\{\cdot\}$ denotes the expectation operator. $\text{i} = \sqrt{-1}$. $x \sim y$ denotes x and y have identical distributions.

[1]Work supported by the National Science Foundation under grant CNS-0905425.

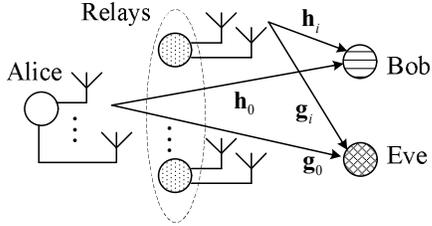

Fig. 1. System model.

## II. SYSTEM AND SIGNAL MODELS

We consider a Gaussian MISO wiretap channel with $N$ helpers, as shown in Fig. 1. The transmitter, Alice, uses $N_t$ antennas to send messages to the legitimate receiver, Bob, through the channel $\mathbf{h}_0^* \in \mathbb{C}^{N_t}$. The eavesdropper, Eve, intercepts messages through the channel $\mathbf{g}_0^* \in \mathbb{C}^{N_t}$. The transmitter is aided by $N$ helpers; each helper has $N_k$ antennas ($N_k \geq 2, k = 1, \cdots, N$) and transmits noise to confound Eve. The channel from helper $k$ to Bob is denoted by $\mathbf{h}_k^* \in \mathbb{C}^{N_k}, k = 1, \cdots, N$, and the channel from helper $k$ to Eve is denoted by $\mathbf{g}_k^* \in \mathbb{C}^{N_k}, k = 1, \cdots, N$.

At Bob or Eve, the received signal is a combination of the source signal, jamming noise and additive white Gaussian noise (AWGN). The received signal at Bob and Eve can be expressed, respectively, as:

$$y_b = \sqrt{P_s}\mathbf{h}_0^\dagger \mathbf{x} + \sum_{k=1}^{N} \mathbf{h}_k^\dagger \mathbf{n}_k + n_b \quad (1)$$

$$y_e = \sqrt{P_s}\mathbf{g}_0^\dagger \mathbf{x} + \sum_{k=1}^{N} \mathbf{g}_k^\dagger \mathbf{n}_k + n_e \quad (2)$$

where $\mathbf{n}_k, k = 1, \cdots, N$ represents the jamming noise that is generated by the helpers; $\mathbf{x}$ is the $N_t \times 1$ source signal vector with input covariance matrix $\mathbf{R}_x \succeq 0$; $n_b$ and $n_e$ are AWGN received at Bob and Eve, respectively, with $\mathbb{E}[|n_b|^2] = \mathbb{E}[|n_e|^2] = N_0$. The source power is constrained as $\text{Tr}(\mathbf{R}_x) \leq P_s$. We may write $\mathbf{R}_x = P_s \mathbf{Q}$, with $\mathbf{Q} \succeq 0$ and $\text{Tr}(\mathbf{Q}) \leq 1$. The power available at helper $k$ is $P_k$.

In the following, we assume that each helper knows only its own link to the destination, $\mathbf{h}_k, k = 1, \cdots, N$, and locally designs its jamming noise so that it delivers a null at Bob.

### A. Nulling Noise Structure

In order for the noise vector of helper $k$ to cause nulling at Bob, it should hold that

$$\mathbf{h}_k^\dagger \mathbf{n}_k = 0, \ k = 1, \ldots, N \quad (3)$$

It is clear that each helper should be equipped with $N_k \geq 2$ antennas, such that there are enough degrees of freedom to design $\mathbf{n}_k$ based on (3). The general solution of (3) can be expressed as $\mathbf{n}_k = \mathbf{E}_k \mathbf{v}_k$, where $\mathbf{E}_k$ ($N_k \times (N_k-1)$) is the null space of $\mathbf{h}_k^\dagger$, with $\mathbf{E}_k^\dagger \mathbf{E}_k = \mathbf{I}$ and $\mathbf{v}_k$ is any arbitrary $(N_k - 1) \times 1$ vector. Let the jamming noise be $\mathbf{n}_k = w_k \mathbf{E}_k \mathbf{t}_k$, where $w_k$ is a weight that will be selected to meet the power constraint, and $\mathbf{t}_k \sim \mathcal{CN}(0, \mathbf{I})$. The power of the jamming noise is

$$\mathbb{E}\{\|\mathbf{n}_k\|^2\} = \mathbb{E}\{\text{Tr}(\mathbf{n}_k \mathbf{n}_k^\dagger)\} = |w_k|^2 \mathbb{E}\{\text{Tr}(\mathbf{E}_k \mathbf{t}_k \mathbf{t}_k^\dagger \mathbf{E}_k^\dagger)\} \quad (4)$$
$$= |w_k|^2 \text{Tr}(\mathbf{E}_k \mathbf{E}_k^\dagger) = (N_k - 1)|w_k|^2.$$

The power constraint is hence $(N_k - 1)|w_k|^2 \leq P_k$.

With this nulling noise, the received signal at Bob and Eve becomes respectively,

$$y_b = \sqrt{P_s}\mathbf{h}_0^\dagger \mathbf{x} + n_b, \quad (5)$$

$$y_e = \sqrt{P_s}\mathbf{g}_0^\dagger \mathbf{x} + \sum_{k=1}^{N} \mathbf{g}_k^\dagger (w_k \mathbf{E}_k \mathbf{t}_k) + n_e. \quad (6)$$

The secrecy rate of this system equals to

$$C_1 = \log_2\left(1 + \frac{P_s}{N_0}\mathbf{h}_0^\dagger \mathbf{Q} \mathbf{h}_0\right)$$
$$- \log_2\left(1 + \frac{P_s \mathbf{g}_0^\dagger \mathbf{Q} \mathbf{g}_0}{\sum_{k=1}^{N} |w_k|^2 \|\mathbf{E}_k^\dagger \mathbf{g}_k\|^2 + N_0}\right). \quad (7)$$

To make $C_1$ as large as possible, the maximum relay power should be used, i.e., it should hold that $(N_k - 1)|w_k|^2 = P_k$. With this, we have

$$C_1 = \log_2\left(1 + \rho_0 \mathbf{h}_0^\dagger \mathbf{Q} \mathbf{h}_0\right)$$
$$- \log_2\left(1 + \frac{\rho_0 \mathbf{g}_0^\dagger \mathbf{Q} \mathbf{g}_0}{\sum_{k=1}^{N} \rho_k(N_k-1)^{-1} \|\mathbf{E}_k^\dagger \mathbf{g}_k\|^2 + 1}\right). \quad (8)$$

where $\rho_0 = P_s/N_0$, $\rho_k = P_k/N_0, k = 1, \cdots, N$ are the Signal-to-Noise Ratio (SNR) at Alice and the $k$th helper.

In the following section we study the problem of maximizing the secrecy rate with respect to $\mathbf{Q}$ using only statistical information about the source-eavesdropper channels.

## III. OUTAGE CONSTRAINED SECRECY RATE MAXIMIZATION

Let us assume that the nodes do not have any channel information with the following exceptions:
- Alice knows $\mathbf{h}_0$ perfectly
- Alice has statistical information on $\mathbf{g}_0$, i.e., $\mathbf{g}_0 \sim \mathcal{CN}(0, \sigma^2 \mathbf{I})$.
- Helper $k$ knows its own link to Bob, $\mathbf{h}_k$.

The outage probability is defined as [14]

$$\text{P}_{\text{out}}(R) = \min_{\mathbf{Q} \succeq 0,\ \text{Tr}(\mathbf{Q}) \leq 1} \Pr(C_1 < R). \quad (9)$$

We will determine the maximum $R$ such that the outage probability is below a prescribed small level, $\epsilon$, which is determined by the qualify of service (QoS) requirements. Mathematically, the problem is formulated as

$$\max_{\mathbf{Q}} R \quad (10)$$
$$\text{s.t.} \quad \text{P}_{\text{out}}(R) \leq \epsilon. \quad (11)$$

The problem of (10) is equivalent to the problem of maximizing the secrecy rate subject to QoS and power constraints as follows

$$\max_{\mathbf{Q}} R \quad (12)$$
$$\text{s.t.} \quad \mathbf{Q} \succeq 0, \quad \text{Tr}(\mathbf{Q}) \leq 1, \quad (13)$$
$$\Pr(C_1 < R) \leq \epsilon. \quad (14)$$

### A. Optimal Input Covariance Matrix Structure

**Lemma 1:** *For the problem of (12), and for $\mathbf{g}_0 \sim \mathcal{CN}(0, \sigma^2 \mathbf{I})$, the optimal $\mathbf{Q}$ is given by $\mathbf{Q}^\star = \mathbf{h}_0 \mathbf{h}_0^\dagger / \|\mathbf{h}_0\|^2$, and the optimization problem can be written as*

$$\max_R R \quad (15)$$

$$\text{s.t. } \Pr\left(\frac{\rho_0 |g_{01}|^2}{\sum_{k=1}^{N} \rho_k (N_k-1)^{-1} \|\mathbf{E}_k^\dagger \mathbf{g}_k\|^2 + 1} > \frac{1 + \rho_0 \|\mathbf{h}_0\|^2}{2^R} - 1\right) \leq \epsilon. \quad (16)$$

The proof is given in Appendix A.

Next, we solve the problem of (15). Obviously, for the optimal $R$, the constraint of (16) holds with equality. Let us define the critical value $\chi_\epsilon$ so that

$$\Pr\Big(\frac{\rho_0|g_{01}|^2}{\sum_{k=1}^N \rho_k(N_k-1)^{-1}\|\mathbf{E}_k^\dagger \mathbf{g}_k\|^2 + 1} > \chi_\epsilon\Big) = \epsilon. \quad (17)$$

Then, the optimal $R$ is given by

$$R^\star = \log(1+\rho_0\|\mathbf{h}_0\|^2) - \log(1+\chi_\epsilon). \quad (18)$$

The condition for $R^\star > 0$ is given in the following lemma.

**Lemma 2:** *For a given $\epsilon$, $R^\star > 0$ if and only if*

$$\Pr\Big(\frac{\rho_0|g_{01}|^2}{\sum_{k=1}^N \rho_k(N_k-1)^{-1}\|\mathbf{E}_k^\dagger \mathbf{g}_k\|^2 + 1} > \rho_0\|\mathbf{h}_0\|^2\Big) < \epsilon. \quad (19)$$

To find the critical value $\chi_\epsilon$, defined in (17), we can use the bisection method. The calculation of the outage probability is discussed the following subsection.

*B. Closed Form Outage Probability*

Let us assume that $\mathbf{g}_0 \sim \mathcal{CN}(0,\sigma^2\mathbf{I})$ and $\mathbf{g}_k \sim \mathcal{CN}(0,\mathbf{\Sigma}_k)$. In this case, the probability of outage can be found in a closed form as follows.

The calculation of the outage probability involves calculating

$$\Pr\Big(\frac{\rho_0|g_{01}|^2}{\sum_{k=1}^N \rho_k(N_k-1)^{-1}\|\mathbf{E}_k^\dagger \mathbf{g}_k\|^2 + 1} > \chi\Big) \quad (20)$$

which can be rewritten as

$$\Pr\Big(\sum_{k=1}^N \rho_k(N_k-1)^{-1}\|\mathbf{E}_k^\dagger \mathbf{g}_k\|^2 - \frac{\rho_0}{\chi}|g_{01}|^2 < -1\Big). \quad (21)$$

$$= \Pr\Big(\frac{\rho_0}{\chi}|g_{01}|^2 - \sum_{k=1}^N \rho_k(N_k-1)^{-1}\|\mathbf{E}_k^\dagger \mathbf{g}_k\|^2 > 1\Big). \quad (22)$$

Note that $\mathbf{E}_k^\dagger \mathbf{g}_k \sim \mathcal{CN}(0, \mathbf{E}_k^\dagger \mathbf{\Sigma}_k \mathbf{E}_k)$. Let $\mathbf{E}_k^\dagger \mathbf{\Sigma}_k \mathbf{E}_k$ have eigen-decomposition $\mathbf{U}_k \mathbf{D}_k \mathbf{U}_k^\dagger$. Denote

$$\mathbf{D} = \mathrm{diag}\Big(\sigma^2\rho_0/\chi, -\frac{\rho_1}{N_1-1}\mathbf{D}_1, \cdots, -\frac{\rho_N}{N_N-1}\mathbf{D}_N\Big). \quad (23)$$

After a few derivations, (22) is equivalent to

$$\Pr(\mathbf{z}^\dagger \mathbf{D} \mathbf{z} > 1) \quad (24)$$

where $\mathbf{z} \sim \mathcal{CN}(0, \mathbf{I}_{M+1})$ with $M = \sum_{k=1}^N (N_k-1)$.

Let $Y = \mathbf{z}^\dagger \mathbf{D} \mathbf{z}$, which is an indefinite quadratic form. The results in [15] give the expression for $\Pr(Y \geq y), y > 0$. Let $\nu_0 = \sigma^2 \rho_0/\chi$ denote the unique positive eigenvalue of $\mathbf{D}$, and $\nu_1, \cdots, \nu_K$ denote different negative diagonal entries of $\mathbf{D}$, with multiplicity $m_1, \cdots, m_K$. Let $m_1 + \cdots + m_K = M$. According to the result in [15, Eq. (32)] for the case $\mathbf{z} \sim \mathcal{CN}(0, \mathbf{I}_{M+1})$, and noting that $\mathbf{D}$ has only one positive eigenvalue, a simple closed form for the outage probability can be obtained as follows:

$$\Pr(Y \geq y) = e^{-y/\nu_0} \prod_{j=1}^K (1-\nu_j/\nu_0)^{-m_j}, \ y > 0 \quad (25)$$

## IV. SIMULATION AND ANALYSIS

In our simulation, Alice has three antennas, and each helper has two antennas. The Alice-Eve link is taken to be $\mathbf{g}_0 \sim \mathcal{CN}(0,\mathbf{I})$, and the $k$th helper-Eve link is taken to be $\mathbf{g}_k \sim \mathcal{CN}(0,\mathbf{I})$. Alice has SNR $\rho_0 = 5$ dB, and all $N$ helpers have SNR $\rho_k = 2$ dB. The outage probability constraint is $\epsilon = 0.01$, and it holds that $\mathrm{P}_{out}(R^*) = \epsilon$, where $R^*$ is given in (18). The results are averaged over $10^5$ independent trials. For each trial, we generate independent and identical distributed $\mathbf{h}_0$, where $\mathbf{h}_0 \sim \mathcal{CN}(0,\mathbf{I})$. For $N = 5, \cdots, 10$ number of helpers, we use bi-section method to search for a larger $R$ that satisfying the outage constraint, until it converges to $R^*$. For a $R^* > 0$, Lemma 2 has to be satisfied. Typically, it is required that the legitimate channel be stronger than the eavesdropping channel, or Alice have a sufficient number of helpers. Fig. 2 shows the obtained secrecy rate as function of the number of helpers. Fig. 3 shows the outage probability for a fixed secrecy rate as function of the number of helpers. Here, we assume that the target secrecy rate is $R_1 = 0.6\log_2(1+P_s\|\mathbf{h}_0\|^2)$. At target secrecy rate $R_1$, the outage probability decreases when the number of helpers increases; this is because Bob is not affected by the jamming noise, and the the more helpers, the more confounded Eve will be.

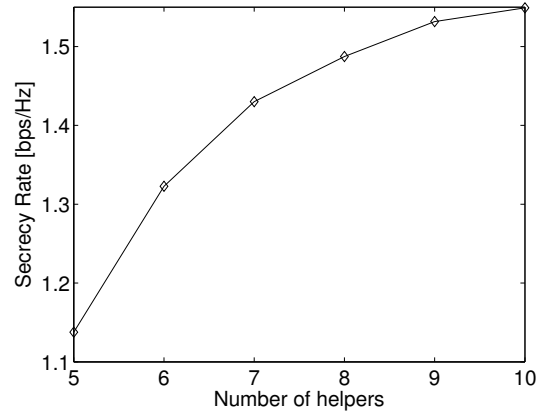

Fig. 2. Secrecy Rate vs Number of Helpers. $\rho_0 = 5$ dB, $\rho_k = 2$ dB. The Outage probability constraint is $\epsilon = 0.01$.

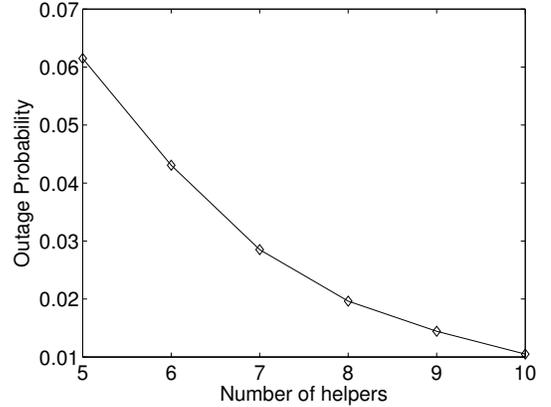

Fig. 3. Outage Probability vs Number of Helpers at a target rate $R = 0.6\log_2(1+P_s\|\mathbf{h}_0\|^2)$. $\rho_0 = 5$ dB, $\rho_k = 2$ dB.

## V. CONCLUSIONS

We have proposed a cooperative jamming scheme, where multiple helpers transmit nulling noise to maximize secrecy rate subject to an outage probability constraint, and a power constraint. Assuming that the transmitter only knows its channel to the legitimate receiver and has statistical CSI on its channel to eavesdropper, and each helper knows only its own link to receiver, we have formulated and solved an outage constrained secrecy rate maximization problem.

## APPENDIX A
## PROOF OF LEMMA 1

After a few derivations we get

$$\Pr(C_1 < R) = \Pr\Big(\frac{\rho_0 \mathbf{g}_0^\dagger \mathbf{Q} \mathbf{g}_0}{\sum_{k=1}^N \rho_k (N_k - 1)^{-1} \|\mathbf{E}_k^\dagger \mathbf{g}_k\|^2 + 1} > \frac{1 + \rho_0 \mathbf{h}_0^\dagger \mathbf{Q} \mathbf{h}_0}{2^R} - 1\Big). \quad (26)$$

To determine the structure of the optimal $\mathbf{Q}$, let $\mathbf{Q}$ have an eigen-decomposition $\mathbf{Q} = \mathbf{U}\mathbf{\Lambda}\mathbf{U}^\dagger$ where $\mathbf{\Lambda} = \mathrm{diag}(\lambda_1, \cdots, \lambda_{N_t})$ is the matrix of eigenvalues with $\lambda_1 \geq \cdots \geq \lambda_{N_t}$, and $\mathbf{U}$ is an unitary matrix. First, we notice that

$$\mathbf{g}_0^\dagger \mathbf{Q} \mathbf{g}_0 = (\mathbf{U}^\dagger \mathbf{g}_0)^\dagger \mathbf{\Lambda}(\mathbf{U}^\dagger \mathbf{g}_0) \sim \mathbf{g}_0^\dagger \mathbf{\Lambda} \mathbf{g}_0 = \sum_{i=1}^{N_t} \lambda_i |g_{0i}|^2 \quad (27)$$

where we have used the fact that if $\mathbf{g}_0 \sim \mathcal{CN}(0, \sigma_0^2 \mathbf{I})$, then $\mathbf{U}^\dagger \mathbf{g}_0 \sim \mathbf{g}_0$. According to (26) and (27), we have

$$\Pr(C_1 < R) = \Pr\Big(\frac{\rho_0 \mathbf{g}_0^\dagger \mathbf{\Lambda} \mathbf{g}_0}{\sum_{k=1}^N \rho_k (N_k - 1)^{-1} \|\mathbf{E}_k^\dagger \mathbf{g}_k\|^2 + 1} > \frac{1 + \rho_0 \mathbf{h}_0^\dagger \mathbf{U}\mathbf{\Lambda}\mathbf{U}^\dagger \mathbf{h}_0}{2^R} - 1\Big). \quad (28)$$

From (28), we know that for the optimal $\mathbf{\Lambda}$, the optimal $\mathbf{U}$ should maximize $\mathbf{h}_0^\dagger \mathbf{U}\mathbf{\Lambda}\mathbf{U}^\dagger \mathbf{h}_0$, since $\frac{1+\rho_0 \mathbf{h}_0^\dagger \mathbf{U}\mathbf{\Lambda}\mathbf{U}^\dagger \mathbf{h}_0}{2^R} - 1$ is increasing with $\mathbf{h}_0^\dagger \mathbf{U}\mathbf{\Lambda}\mathbf{U}^\dagger \mathbf{h}_0$ but decreasing with $R$, in other words, a larger $\mathbf{h}_0^\dagger \mathbf{U}\mathbf{\Lambda}\mathbf{U}^\dagger \mathbf{h}_0$ will allow a larger $R$ without violating the outage constraint $\Pr(C_1 < R) \leq \epsilon$.

Let $\mathbf{U}^\dagger \mathbf{h}_0 = \mathbf{y} = [y_1, \cdots, y_{N_t}]^T$. Then $\|\mathbf{y}\|^2 = \|\mathbf{h}_0\|^2$. With this, we write

$$\mathbf{h}_0^\dagger \mathbf{U}\mathbf{\Lambda}\mathbf{U}^\dagger \mathbf{h}_0 = \mathbf{y}^\dagger \mathbf{\Lambda} \mathbf{y} \leq \lambda_1 \|\mathbf{y}\|^2 = \lambda_1 \|\mathbf{h}_0\|^2. \quad (29)$$

Equality holds if $\mathbf{y} = [\|\mathbf{h}_0\|, 0, \cdots, 0]^T$. It follows from $\mathbf{U}^\dagger \mathbf{h}_0 = \mathbf{y}$ that $\mathbf{U} = [\mathbf{h}_0/\|\mathbf{h}_0\|, \mathbf{u}_2, \cdots, \mathbf{u}_{N_t}]$. According to (27), (28) and (29), we have

$$\Pr(C_1 < R) = \Pr\Big(\frac{\rho_0 \sum_{i=1}^{N_t} \lambda_i |g_{0i}|^2}{\sum_{k=1}^N \rho_k (N_k - 1)^{-1} \|\mathbf{E}_k^\dagger \mathbf{g}_k\|^2 + 1} > \frac{1 + \rho_0 \lambda_1 \|\mathbf{h}_0\|^2}{2^R} - 1\Big). \quad (30)$$

From (30), we know that for the optimal $\lambda_1$, smaller values of $\lambda_2, \cdots, \lambda_{N_t}$ will allow a larger $R$ without violating the outage constraint $\Pr(C_1 < R) \leq \epsilon$. Thus, it should hold that $\lambda_2 = \cdots = \lambda_{N_t} = 0$. As a result, we have

$$\Pr(C_1 < R) = \Pr\Big(\frac{\rho_0 |g_{01}|^2}{\sum_{k=1}^N \rho_k (N_k - 1)^{-1} \|\mathbf{E}_k^\dagger \mathbf{g}_k\|^2 + 1} > \frac{\rho_0 \|\mathbf{h}_0\|^2}{2^R} - \frac{1 - \frac{1}{2^R}}{\lambda_1}\Big). \quad (31)$$

From (31), we know that $\frac{\rho_0 \|\mathbf{h}_0\|^2}{2^R} - \frac{1-\frac{1}{2^R}}{\lambda_1}$ is increasing with $\lambda_1$ but decreasing with $R$, in other words, a larger $\lambda_1$ will allow a larger $R$ without violating the outage constraint $\Pr(C_1 < R) \leq \epsilon$. Since $\mathrm{Tr}(\mathbf{Q}) = \lambda_1 \leq 1$, it holds that $\lambda_1 = 1$. As a result, we have

$$\Pr(C_1 < R) = \Pr\Big(\frac{\rho_0 |g_{01}|^2}{\sum_{k=1}^N \rho_k (N_k - 1)^{-1} \|\mathbf{E}_k^\dagger \mathbf{g}_k\|^2 + 1} > \frac{1 + \rho_0 \|\mathbf{h}_0\|^2}{2^R} - 1\Big). \quad (32)$$

Based on the result above, the problem of (12) is equivalent to the problem of (15). This completes the proof.